\documentclass[aps,preprint,showpacs,amssymb,groupedaddress]{revtex4}
\usepackage{url,hyperref,lineno,microtype}
\usepackage{bm}
\usepackage{graphicx}
\usepackage{mathrsfs}
\usepackage{amsmath}
\usepackage{epstopdf}

\newcommand{\ppdfew}{\ensuremath{{\rm PPD}_{A=2}}}
\newcommand{\ppdmany}{\ensuremath{{\rm PPD}_{A=2,3,4}}}

\def\nuc#1#2{\relax\ifmmode{}^{#1}{\protect\text{#2}}\else${}^{#1}$#2\fi}
\def\itnuc#1#2{\setbox\@tempboxa=\hbox{\scriptsize\it #1}
  \def\@tempa{{}^{\box\@tempboxa}\!\protect\text{\it #2}}\relax
  \ifmmode \@tempa \else $\@tempa$\fi}

\newcommand{\newabbreviation}[3]{\newcounter{#1}\expandafter\newcommand\csname#1\endcsname[1][]{\ifthenelse{\equal{##1}{abreviate}}{#2}{\ifthenelse{\equal{##1}{fullname}}{#3}{\ifthenelse{\equal{##1}{explain}}{#3 (#2)\stepcounter{#1}}{\ifthenelse{\value{#1}=0}{#3##1 (#2##1)\stepcounter{#1}}{#2##1}}}}}}

\newabbreviation{EFT}{EFT}{effective field theory}
\newabbreviation{chiEFT}{$\chi$EFT}{chiral effective field theory}
\newabbreviation{LO}{\text{LO}}{leading order}
\newabbreviation{NLO}{\text{NLO}}{next-to-leading order}
\newabbreviation{NNLO}{\text{NNLO}}{next-to-next-to-leading order}
\newabbreviation{LEC}{\text{LEC}}{low-energy constant}

\newabbreviation{EC}{\text{EC}}{eigenvector continuation}
\newabbreviation{CC}{\text{CC}}{coupled-cluster}
\newabbreviation{NCSM}{\text{NCSM}}{no-core shell model}
\newabbreviation{SPCC}{\text{SPCC}}{subspace-projected coupled cluster}
\newabbreviation{PPD}{\text{PPD}}{posterior predictive distribution}
\newabbreviation{PDF}{\text{PDF}}{probability density function}
\newabbreviation{NM}{\text{NM}}{nuclear matter}
\newabbreviation{PNM}{\text{PNM}}{pure neutron matter}
\newabbreviation{SNM}{\text{SNM}}{symmetric nuclear matter}
\newabbreviation{EOS}{\text{EOS}}{equation-of-state}
\newabbreviation{MCMC}{\text{MCMC}}{Markov chain Monte Carlo}
\newabbreviation{SIR}{\text{S/IR}}{sampling/importance resampling}

\DeclareUnicodeCharacter{202F}{\,}

\usepackage{todonotes}

\begin{document}

\title[Bayesian probability updates]{Bayesian probability updates using Sampling/Importance Resampling: Applications in nuclear theory} 
\author{W.\ G.\ Jiang}
\affiliation{Department of Physics, Chalmers University of Technology, SE-412 96 G\"oteborg, Sweden}

\author{C.\ Forss\'en}
\affiliation{Department of Physics, Chalmers University of Technology, SE-412 96 G\"oteborg, Sweden}

\begin{abstract}
We review an established Bayesian sampling method called sampling/importance resampling and highlight situations in nuclear theory when it can be particularly useful. To this end we both analyse a toy problem and demonstrate realistic applications of importance resampling to infer the posterior distribution for parameters of $\Delta$NNLO interaction model based on chiral effective field theory and to estimate the posterior probability distribution of target observables. The limitation of the method is also showcased in extreme situations where importance resampling breaks.

\end{abstract}


\maketitle

\section{Introduction}
Bayesian inference is an appealing approach for dealing with theoretical uncertainties and has been applied in different nuclear physics studies~\cite{schindler2009,caesar2013,furnstahl2015,wesolowski2019,melendez2019,Epelbaum:2019zqc,Yang:2020,Phillips:2020,drischler2020,drischler2020b,Maris:2020qne,Wesolowski:2021cni,Djarv:2021pjc,Svensson:2021lzs}. In the practice of Bayesian analyses, a sampling procedure is usually inevitable for approximating the posterior probability distribution of model parameters and for performing predictive computations. Various \MCMC{} methods~\cite{Metropolis:1953,Hastings:1970,Hitchcock:2003,Toussaint2011,brooks2011handbook} are often used for this purpose, even for complicated models with high-dimensional parameter spaces. However, \MCMC{} sampling typically requires many likelihood evaluations, which is often a costly operation in nuclear theory, and there is a need to explore other sampling techniques. In this paper, we review an established method called \SIR{}~\cite{Rubin:1988, Smith:1992aa, Bernardo:2006} and demonstrate its use in realistic nuclear physics applications where we also perform comparisons with \MCMC{} sampling.

In recent years, there has been an increasing demand for precision nuclear theory.This implies a challenge to not just achieve accurate theoretical predictions but also to quantify accompanying uncertainties. The use of \emph{ab initio} many-body methods and nuclear interaction models based on \chiEFT{} has shown a potential to describe finite nuclei and nuclear matter based on extant experimental data (e.g. nucleon-nucleon scattering, few-body sector) with controlled approximations \cite{vankolck1999,bogner2003,epelbaum2009,bogner2010,machleidt2011}. The interaction model is parametrized in terms of \LEC[s], the number of which is growing order-by-order according to the rules of a corresponding power counting \cite{weinberg1990,weinberg1991,kaplan1998}. Very importantly, the systematic expansion allows to quantify the truncation error and to incorporate this knowledge in the analysis~\cite{wesolowski2019,melendez2019,Epelbaum:2019zqc,drischler2020b,Maris:2020qne,Wesolowski:2021cni,Djarv:2021pjc,Svensson:2021lzs}. Indeed, Bayesian inference is an excellent framework to incorporate different sources of uncertainty and to propagate error bars to the model predictions. Starting from Bayes' theorem
\begin{equation}\label{eq:posterior_hyperparameter}
\rm{pr}(\bm{\theta}|\mathcal{D}) \propto \mathcal{L}(\bm{\theta})\rm{pr}(\bm{\theta}),
\end{equation}
where $\rm{pr}(\bm{\theta}|\mathcal{D})$ is the posterior \PDF{} for the vector $\bm{\theta}$ of \LEC[s] (conditional on the data $\mathcal{D}$), $\mathcal{L}(\bm{\theta}) \equiv \rm{pr}(\mathcal{D}~|\bm{\theta})$ is the likelihood and $\rm{pr}(\bm{\theta})$ is the prior. Then for any model prediction one needs to evaluate the expectation value of a function of interest $\bf{y}(\bm{\theta})$ (target observables) according to the posterior. This involves integrals such as
\begin{equation}\label{eq:ppd_integral}
  \int d\bm{\theta} \bm{y}(\bm{\theta}) \rm{pr}(\bm{\theta}|\mathcal{D}),
\end{equation}
which can not be analytically solved for realistic cases. Fortunately, integrals such as Eq.~\eqref{eq:ppd_integral} can be approximately evaluated using a finite set of samples $\{ \bm{\theta}_i \}_{i=1}^N$ from $\rm{pr}(\bm{\theta}|\mathcal{D})$. \MCMC{} sampling methods are the main computational tool for providing such samples, even for high-dimensional parameter volumes~\cite{Svensson:2022kkj}. However the use of \MCMC{} in nuclear theory typically requires massive computations to record sufficiently many samples from the Markov chain. There are certainly situations where \MCMC{} sampling is not ideal, or even becomes infeasible:
\begin{enumerate}
\item{When the posterior is conditioned on some calibration data for which our model evaluations are very costly. Then we might only afford a limited number of full likelihood evaluations and our \MCMC[] sampling becomes less likely to converge.}
\item{Bayesian posterior updates in which calibration data is added in several different stages. This typically requires that the \MCMC[] sampling must be carried out repeatedly from scratch.}
\item{Model checking where we want to explore the sensitivity to prior assignments. This is a second example of posterior updating}.
  \item{The prediction of target observables for which our model evaluations become very costly and the handling of a large number of \MCMC{} samples becomes infeasible.}
\end{enumerate}

These are situations where one might want to use the \SIR{} method~\cite{Smith:1992aa,Bernardo:2006}, which allows posterior probability updates with a minimum amount of computation (previous results of model evaluations remain useful). In the following sections we first review the \SIR{} method and then present both toy and realistic applications in which its performance is compared with full \MCMC{} sampling. Finally, we illustrate limitations of the method by considering cases where \SIR{} fails and we highlight the importance of the so-called effective number of samples. More difficult scenarios, in which the method fails without a clear warning, are left for the concluding remarks.

\section{Sampling/Importance resampling}
The basic idea of \SIR{} is to utilize the inherent duality between samples and the density (probability distribution) from which they were generated~\cite{Smith:1992aa}. This duality offers an opportunity to indirectly recreate a density (that might be hard to compute) from samples that are easy to obtain. Here we give a brief review of the method and illustrate with a toy problem.

Let us consider a target density $h(\bm{\theta})$. In our applications this target will be the posterior \PDF{} $\rm{pr}(\bm{\theta}|\mathcal{D})$ from Eq.~\eqref{eq:posterior_hyperparameter}. Instead of attempting to directly collect samples from $h(\bm{\theta})$, as would be the goal in \MCMC{} approaches, the \SIR{} method uses a detour. We first obtain samples from a simple (even analytic) density $g(\bm{\theta})$. We then resample from this finite set using a resampling algorithm to approximately recreate samples from the target density $h(\bm{\theta})$. There are (at least) two different resampling methods. In this paper we only focus on one of them called weighted bootstrap (more details of resampling methods can be found in Refs.~\cite{Rubin:1988, Smith:1992aa}).

Assuming we are interested in the target density $h(\bm{\theta})=f(\bm{\theta})\,/\int\! f(\bm{\theta})\,\mathrm{d}\bm{\theta}$, the procedure of resampling via weighted bootstrap can be summarized as follows:
\begin{enumerate}
  \item Generate the set $\{ \bm{\theta}_i\}_{i=1}^n$ of samples from a sampling density $g(\bm{\theta})$.
  \item Calculate $\omega_i=f(\bm{\theta}_i)\,/\,g(\bm{\theta}_i)$ for the $n$ samples and define importance weights as: $q_i=\omega_i~/\sum_{j=1}^n \omega_j$.
  \item Draw $N$ new samples $\{ \bm{\theta}_i^*\}_{i=1}^N$ from the discrete distribution $\{ \bm{\theta}_i \}_{i=1}^n$ with probability mass $q_i$ on $\bm{\theta}_i$.
  \item The set of samples $\{ \bm{\theta}_i^* \}_{i=1}^N $ will then be approximately distributed according to the target density $h(\bm{\theta})$.
\end{enumerate}

Intuitively, the distribution of $\bm{\theta}^*$ should be good approximation of $h(\bm{\theta})$ when $n$ is large enough. Here we justify this claim via the cumulative distribution function of $\theta^*$ (for the one-dimensional case)
\begin{equation}
\begin{aligned}
\rm{pr}(\theta^*\leq a) &= \sum\limits_{i=1}^n q_i \cdot H(a-\theta_i) 
= \frac{ \frac{1}{n}\sum\limits_{i=1}^n \omega_i \cdot H(a-\theta_i)}{ \frac{1}{n}\sum\limits_{i=1}^n \omega_i} \\
& \xrightarrow[n \rightarrow \infty]{} \frac{\mathbb{E}_g\left[ \frac{f(\theta)}{g(\theta)} \cdot H(a-\theta_i) \right]}{\mathbb{E}_g\left[\frac{f(\theta)}{g(\theta)}\right]}
= \frac{\int^{a}_{-\infty}f(\theta)\,d\theta}{\int^{\infty}_{-\infty}f(\theta)\,d\theta}= \int^{a}_{-\infty}h(\theta)\,d\theta ,
\end{aligned} 
\label{eq:cdf}
\end{equation}
with $\mathbb{E}_g[X(\theta)]=\int^{\infty}_{-\infty} X(\theta) g(\theta)\,d\theta$ the expectation value of $X(\theta)$ with respect to $g(\theta)$, and $H$ Heaviside step function such that
\begin{equation}
   H(a-\theta_i) = \begin{cases}
      1 & \text{if}\ \theta_i \leq a, \\
      0 & \text{if}\ \theta_i > a .
\end{cases}
\label{eq:Heaviside}
\end{equation}

The above resampling method can be applied to generate samples from the posterior \PDF{} $h(\bm{\theta})=\rm{pr}(\bm{\theta}|\mathcal{D})$ in a Bayesian analysis. It remains to choose a sampling distribution, $g(\bm{\theta})$, which in principle could be any continuous density distribution. However, recall that $h(\bm{\theta})$ can be expressed in terms of an unnormalized distribution $f(\bm{\theta})$, and using Bayes' theorem~\eqref{eq:posterior_hyperparameter} we can set $f(\bm{\theta})=\mathcal{L}(\bm{\theta})\rm{pr}(\bm{\theta})$. Thus, choosing the prior $\rm{pr}(\bm{\theta})$ as the sampling distribution $g(\bm{\theta})$ we find that the importance weights are expressed in terms of the likelihood, $q_i = \mathcal{L}(\bm{\theta}_i)/ \sum_{j=1}^n \mathcal{L}(\bm{\theta}_j)$. Assuming that it is simple to collect samples from the prior, the costly operation will be the evaluation of $\mathcal{L}(\bm{\theta}_i)$. Here we make the side remark that an effective and computationally cost-saving approximation can be made if we manage to perform a pre-screening to identify (and ignore) samples that will give a very small importance weight. We also note that the above choice of $g(\bm{\theta})=\rm{pr}(\bm{\theta})$ is purely for simplicity and one can perform importance resampling with any $g(\bm{\theta})$.

\begin{figure*}[htbp]
  \includegraphics[width=0.95\textwidth] {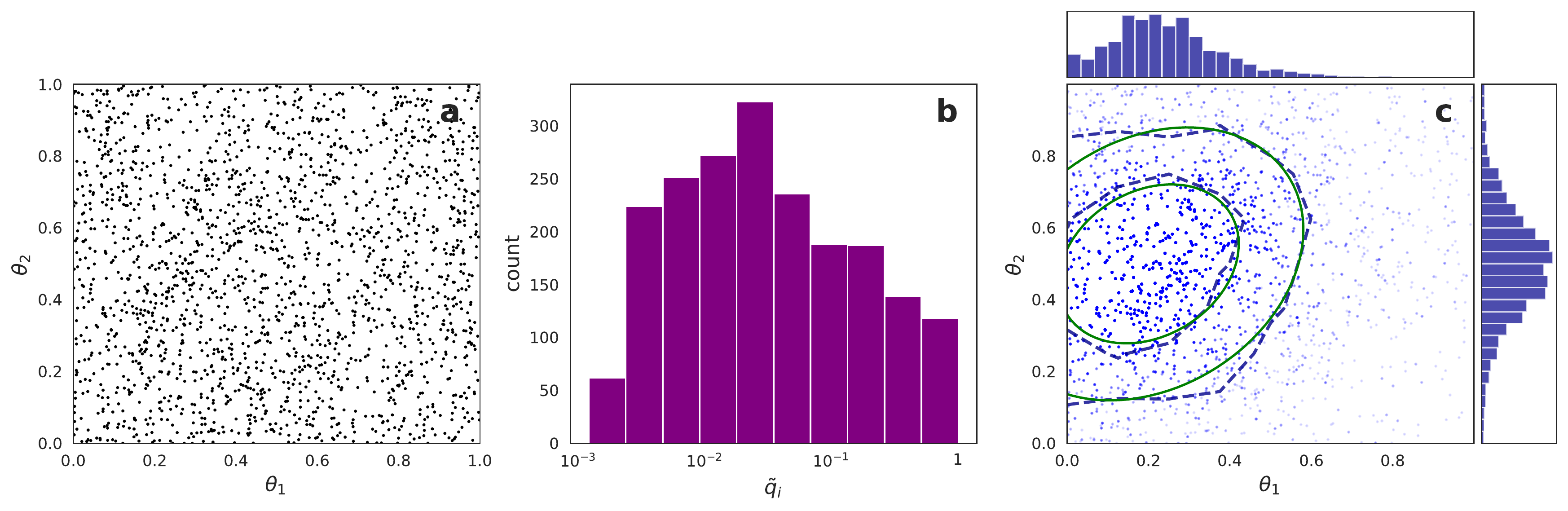} 
  \caption{Illustration of \SIR{} procedures. \textbf{a}. Samples $\{ \bm{\theta}\}_{i=1}^n$ from the uniform prior in a unit square ($n=2000$ samples are shown). \textbf{b}. Histogram of rescaled importance weights $\tilde{q}_i=q_i/\max(\{q\})$ where $q_i= \mathcal{L}(\bm{\theta}_i)/ \sum_{j=1}^n \mathcal{L}(\bm{\theta}_j)$ with $\mathcal{L}(\bm{\theta})$ as in Eq.~\eqref{eq:student-t}. The number of effective samples is $n_\mathrm{eff}=214.6$. Note that the samples are drawn from a unit square and that the tail of the target distribution is not covered. \textbf{c}. Samples $\{ \bm{\theta}^*\}_{i=1}^N$ of the posterior (blue dots with 10\% opacity) resampled from the prior samples with probability mass $q_i$. The contour lines for the $68\%$ and $90\%$ credible regions of the posterior samples (blue dashed) are shown and compared with those of the exact bivariate target distribution (green solid). Summary histograms of the marginal distributions for $\theta_1$ and $\theta_2$ are shown in the top and right subplots.
  \label{fig:SIR}}
\end{figure*}

In Fig.~\ref{fig:SIR} we follow the above procedure and give a simple example of \SIR{} to illustrate how to get samples from a posterior distribution. We consider a two-dimensional parametric model with $\bm{\theta} = (\theta_1$, $\theta_2)$. Given data $\mathcal{D}$ obtained under the model we have:
\begin{equation}
\rm{pr}(\theta_1,\theta_2|\mathcal{D})=\frac{ \mathcal{L}(\theta_1,\theta_2)\rm{pr}(\theta_1,\theta_2)}{\iint \mathcal{L}(\theta_1,\theta_2) \rm{pr}(\theta_1,\theta_2) \,d\theta_1d\theta_2}.
\label{eq:example_1}
\end{equation}
For simplicity and illustration, the joint prior distribution for $\theta_1$, $\theta_2$ is set to be uniform over the unit square as shown in Fig.~\ref{fig:SIR}a. In this example we also assume that the data $\mathcal{D}$ follows a multivariate Student-t distribution such that the likelihood function is
\begin{equation}
  \mathcal{L}(\theta_1,\theta_2) = 
\frac{\Gamma[(\nu+p)/2]}{\Gamma(\nu/2)\nu^{p/2}\pi^{p/2}|\bm{\Sigma}|^{1/2}}\left[ {1+\frac{1}{\nu}(\bm{\theta}-\bm{\mu})^{T}\bm{\Sigma}^{-1}(\bm{\theta}-\bm{\mu})} \right]^{-(\mu+p)/2},
\label{eq:student-t}
\end{equation}
where the dimension $p=2$, the degrees of freedom $\nu=2$, the mean vector $\bm{\mu} = (0.2, 0.5)$ and the scale matrix $\bm\Sigma=[[0.02, 0.005], [0.005, 0.02]]$.

The importance weights $q_i$ are then computed for $n=2000$ samples drawn from the prior (these prior samples are shown in Fig.~\ref{fig:SIR}a). The resulting histogram of importance weights is shown in Fig.~\ref{fig:SIR}b. Here the weights have been rescaled as $\tilde{q}_i=q_i/\max(\{ q \})$ such that the sample with the largest probability mass corresponds to 1 in the histogram. We also define the effective number of samples, $n_\mathrm{eff}$, as the sum of rescaled importance weights, $n_\mathrm{eff} = \sum_{i=1}^n \tilde{q}_i$. Finally, in Fig.~\ref{fig:SIR}c we show $N=20,000$ new samples $\{ \bm{\theta}_i^* \}_{i=1}^N$ that are drawn from the prior samples $\{ \bm{\theta}_i\}_{i=1}^n$ according to the probability mass $q_i$ for each $\bm{\theta}_i$. The blue and green contour lines represent (68\% and 90\%) credible regions for the resampled distribution and for the Student-t distribution, respectively. This result demonstrates that the samples generated by the \SIR{} method give a very good approximation of the target posterior distribution.

\section{Nuclear physics applications}
Now that we have reviewed the basic idea of the \SIR{} method, we move on to present realistic applications of the resampling technique in nuclear structure calculations. Here we study Bayesian inference involving the $\Delta$NNLO chiral interaction~\cite{ekstrom2017} with explicit inclusion of delta isobar degree of freedom at next-to-next-to-leading order. In Weinberg's power counting the $\Delta$NNLO interaction model is parametrized by 17 \LEC[s], with four pion-nucleon \LEC[s] ($c_{1,2,3,4}$) that are inferred from pion-nucleon scattering data and 13 additional \LEC[s] that should be inferred from extant experimental data of low-energy nucleon-nucleon scattering and bound-state nuclear observables. 

For this application we treat only a subset of the parameters as active and keep the other \LEC[s] fixed at values taken from the $\Delta \rm{NNLO_{GO}}(450)$ interaction~\cite{jiang2020}. Specifically, we consider deuteron observables and use seven active model parameters: $c_{1,2,3,4}$, $\tilde{C}_{3S1}$, $C_{3S1}$, $C_{E1}$. Our Gaussian likelihood contains three independent data: the deuteron ground state energy $E$, its point-proton radius $R_\mathrm{p}$ and quadrupole moment $Q$ with experimental targets from Refs.~\cite{Wang:2020,angeli2013} (for $Q$ we use the theoretical result obtained by the CD-Bonn\cite{machleidt2001} model). With these simplified conditions, we perform \SIR{} as well as \MCMC{} sampling to study (1) the posterior \PDF{} for the LECs and (2) \PPD[s] for selected few-body observables. This application therefore allows a straightforward comparison of the two different sampling methods in a realistic setting. 

It is the computation of observables, e.g., for likelihood evaluations, which is usually the major, time-consuming bottleneck in Bayesian analyses using \MCMC{} methods. In this application, the statistical analysis is enabled by the use of emulators which mimic the outputs of many-body solvers but are faster by orders of magnitude. The emulators employed here for the ground-state observables of the deuteron, and later for few-body observables, are based on eigenvector continuation~\cite{frame2018,Konig:2019adq,ekstrom2019}. These emulators allow to reduce the computation time from seconds to milliseconds while keeping the relative error (compared with full no-core shell model calculation) within $0.001\%$. Unfortunately, emulators are not yet available for all nuclear observables. The \MCMC{} sampling of posterior \PDF[s], or the evaluation of expectation integrals such as Eq.~\eqref{eq:ppd_integral}, will typically not work for models with observables that require heavy calculations.

\begin{table}[!htb]
 \caption
 {Experimental target values, $z$, and error assignments, $\varepsilon$, for observables used in the model calibration and for predictions. Energies in [MeV], point-proton radii in [fm], and the deuteron quadrupole moment in [$e^2\mathrm{fm}^2$]. 
}
\label{tab:error_assignments}
\begin{center}
   \begin{tabular}{cccccc}
   \hline
   \multicolumn{6}{c}{Calibration observables}\\
   \hline
    Observable  & $z$ & $\varepsilon_\mathrm{exp}$ & $\varepsilon_\mathrm{model}$
    & $\varepsilon_\mathrm{method}$ & $\varepsilon_\mathrm{em}$    \\
    \hline 
    $E(^{2}\mathrm{H})$ &  -2.2298   &     0    & 0.05 & 0.0005     & 0.001\%  \\
    %
    $R_p(^{2}\mathrm{H})$ & 1.976    &    0     & 0.005     & 0.0002 &0.0005\%   \\
    $Q(^{2}\mathrm{H})$ & 0.27      &      0.01   & 0.003 &   0.0005     &0.001\%   \\
   \hline
   \multicolumn{6}{c}{Predicted observables}\\
   \hline
    $E(^{3}\mathrm{H})$ &  -8.4818   &     0	  & 0.17	 & 0.0005         &0.01\%  \\
    $E(^{4}\mathrm{He})$ & -28.2956   &   0     & 0.55     &   0.0005       &0.01\%  \\
    $R_\mathrm{p}(^{4}\mathrm{He})$ & 1.455  &  0   & 0.016   &  0.0002  &0.003\%  \\
    \hline
  \end{tabular}
\end{center}
\end{table}

\begin{figure}
  \includegraphics[width=0.95\columnwidth] {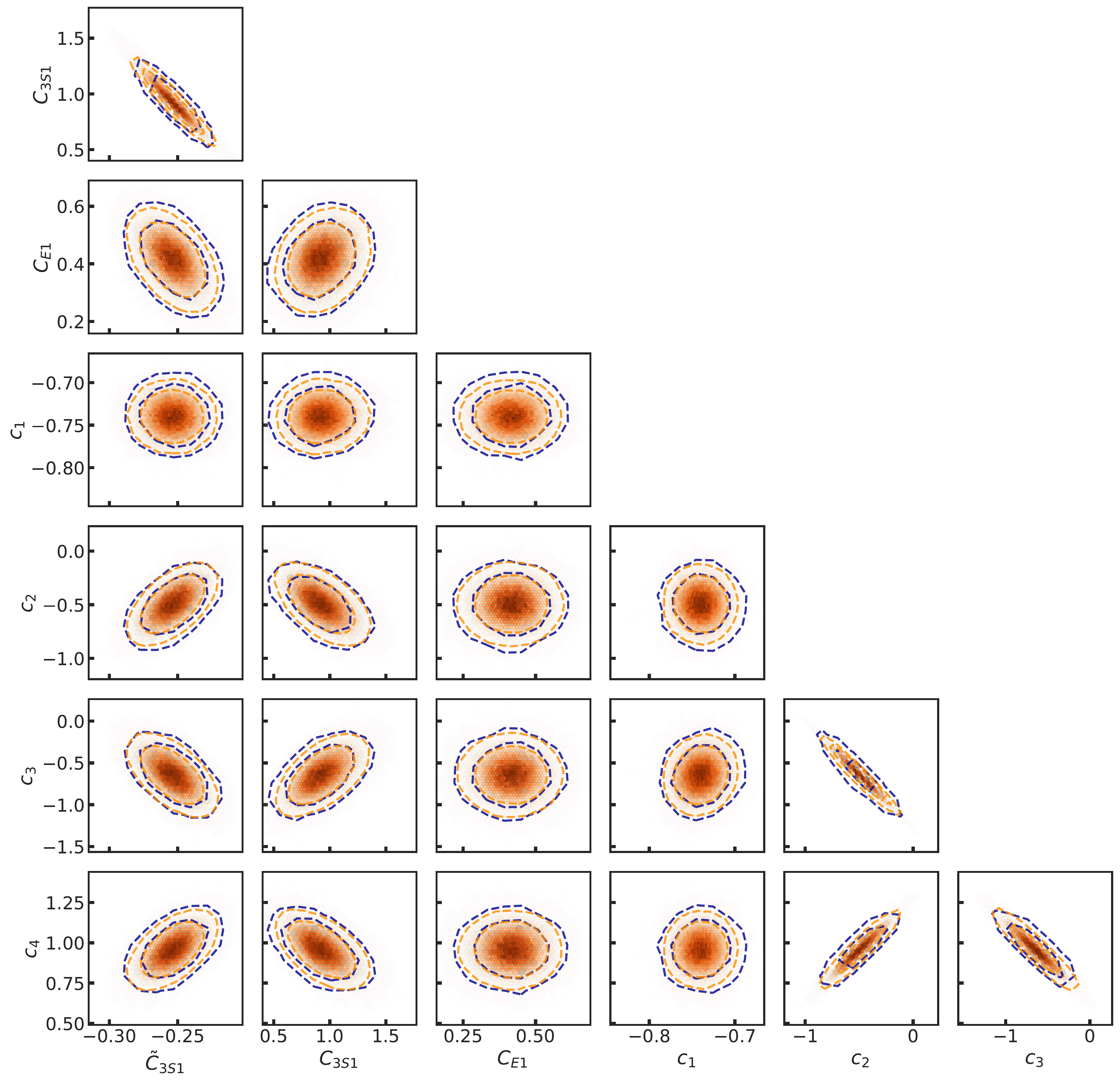}  
  \caption{The joint posterior of LECs sampled with \SIR{} (blue) compared with \MCMC{} sampling (orange). The LECs are shown in units of $10^4\, \rm{GeV^{-1}}$,$10^4\, \rm{GeV^{-2}}$ and $10^4\, \rm{GeV^{-4}}$ for $c_{i}$, $\tilde{C}_i$ and $C_i$, respectively. The likelihood observables and assigned errors are given in Table~\ref{tab:error_assignments}. The contour lines indicate 68\% and 90\% credible regions.
\label{fig:PDF}}
\end{figure}

The experimental target values and error assignments for the calibration observables used to condition the posterior \PDF{} are listed in the upper half of Table~\ref{tab:error_assignments}. In this study we assume a normally-distributed likelihood, and consider different sources of error when calibrating the model predictions with experimental data. The errors are assumed to be independent. They include experimental, $\varepsilon_\mathrm{exp}$, model (\chiEFT{} truncation) discrepancy, $\varepsilon_\mathrm{model}$, many-body method, $\varepsilon_\mathrm{method}$, and emulator, $\varepsilon_\mathrm{em}$, errors. More details on the determination of the error scales can be found in Ref.~\cite{Hu:2021trw}.

\begin{figure}[htbp]
  \includegraphics[width=0.90\textwidth] {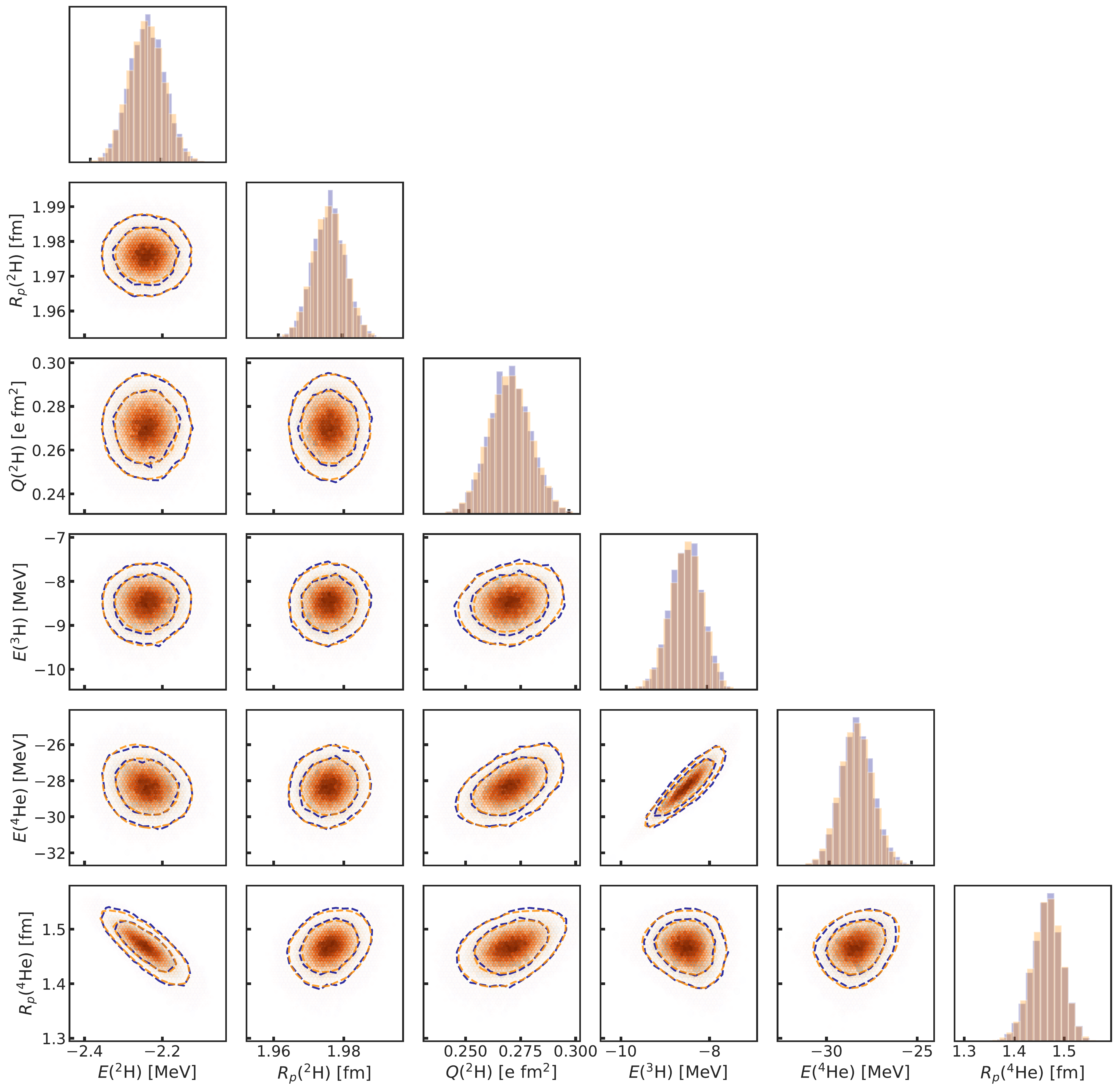}
  \caption{The \PPD{} obtained from samples of the \LEC[s] posterior distribution as shown in Fig.~\ref{fig:PDF}. The bivariate histograms and the corresponding contour lines denote the joint distribution of observables generate by \SIR{} (blue) and \MCMC{} sampling (orange). The marginal distributions of the observables are shown in the diagonal panels.
 \label{fig:PPD}}
\end{figure}

Furthermore, we take advantage of previous studies and incorporate information about $c_{1,2,3,4}$ from a Roy-Steiner analysis of pion-nucleon scattering data~\cite{Siemens:2017} and identify a non-implausible domain for $\tilde{C}_{3S1}$, $C_{3S1}$, $C_{E1}$ from a history matching approach in Ref.~\cite{Hu:2021trw}\footnote{Specifically we use the non-implausible domain that was identified in wave 2 of the history matching performed in Ref.~\cite{Hu:2021trw}. This wave only included deuteron observables.}. With this prior knowledge we set up the prior distribution of the seven \LEC[s] as the product of a multivariate Gaussian for $c_{1,2,3,4}$ and a uniform distribution for $\tilde{C}_{3S1}$, $C_{3S1}$, $C_{E1}$. We note that the use of history matching is very beneficial for both \SIR{} and \MCMC{} sampling. For \SIR{} it allows to select a sampling distribution that promises a large overlap with the target distribution and it identifies prior samples that are likely to have large weights in the resampling step. For \MCMC{}, the non-implausible samples from history matching serve as good starting points for the walkers and thereby give faster convergence.

\subsection{Posterior sampling}
Once we have the prior and the likelihood function we are able to draw samples from the posterior \PDF{} and to analyze the $ab$ $initio$ description of few-nucleon systems with the present interaction model. The joint posterior of the \LEC[s] is shown in Fig.~\ref{fig:PDF}, where we compare bivariate, marginal distributions from \SIR{} and \MCMC{} sampling.
For the \MCMC{} sampling we employed an open-source Python toolkit called emcee~\cite{Foreman_Mackey_2013} that performs affine-invariant ensemble sampling. We use 150 walkers that are warmed up with 5000 initial steps and then move for $5\times10^{5}$ steps. This amounts to $7.6\times10^{7}$ likelihood evaluations. The positions of the walkers are recorded every 500 steps which gives $1.5\times10^{5}$ samples from the posterior distribution of the \LEC[s].
On the other hand, for \SIR{} we first acquire $2\times 10^{4}$ samples from the prior distribution and perform the same number of likelihood evaluations to get the importance weights. From this limited set we then draw $1.5\times10^{5}$ samples using resampling (the same final number as in \MCMC{}). Note that several prior samples occur more than once in the final sample set. Here the number of effective samples for \SIR{} is $n_\mathrm{eff} = 1589.9$. As we can see from Fig.~\ref{fig:PDF}, the contour lines of both sampling methods are in good agreement and, e.g., the correlation structure of the \LEC{} pairs are equally well described. The histograms of \SIR{} and \MCMC{} samples are both plotted in the figure but are almost impossible to distinguish.

As a second stage we employ the inferred model to perform model checking of the calibration observables and to predict the \nuc{3}{H} ground-state energy and the \nuc{4}{He} ground-state energy and point-proton radius (see Table~\ref{tab:error_assignments}). For this purpose the \PPD{} is defined as the set
\begin{equation}
\{ \rm \textbf{y}_{th}(\bm{\theta}) : \bm{\theta} \sim {\rm pr}(\bm{\theta} \, | \, {\mathcal{D}})  \},
\label{eq:ppd}
\end{equation}
where $\rm \textbf{y}_{th}(\bm{\theta})$ is the theoretical predictions of selected observables using the model parameter vector $\bm{\theta}$. Fig.~\ref{fig:PPD} illustrate the \PPD{} of the three deuteron observables using \SIR{} (blue) and \MCMC{} sampling (orange). 
The marginal histograms of the observable predictions are shown in the diagonal panels of the corner plot. In this study both sampling methods give very similar distributions for all observables. Note that the predictive distributions for the three deuteron observables can be considered as model checking since they appeared in the likelihood function and therefore conditioned the \LEC{} posterior. The $^{3}\mathrm{H}$ and $^{4}\mathrm{He}$ observables, on the other hand, are predictions in this study. Their distributions are characterized by larger variances compared to the deuteron predictions. 

\begin{figure}[htbp]
  \includegraphics[width=0.90\textwidth] {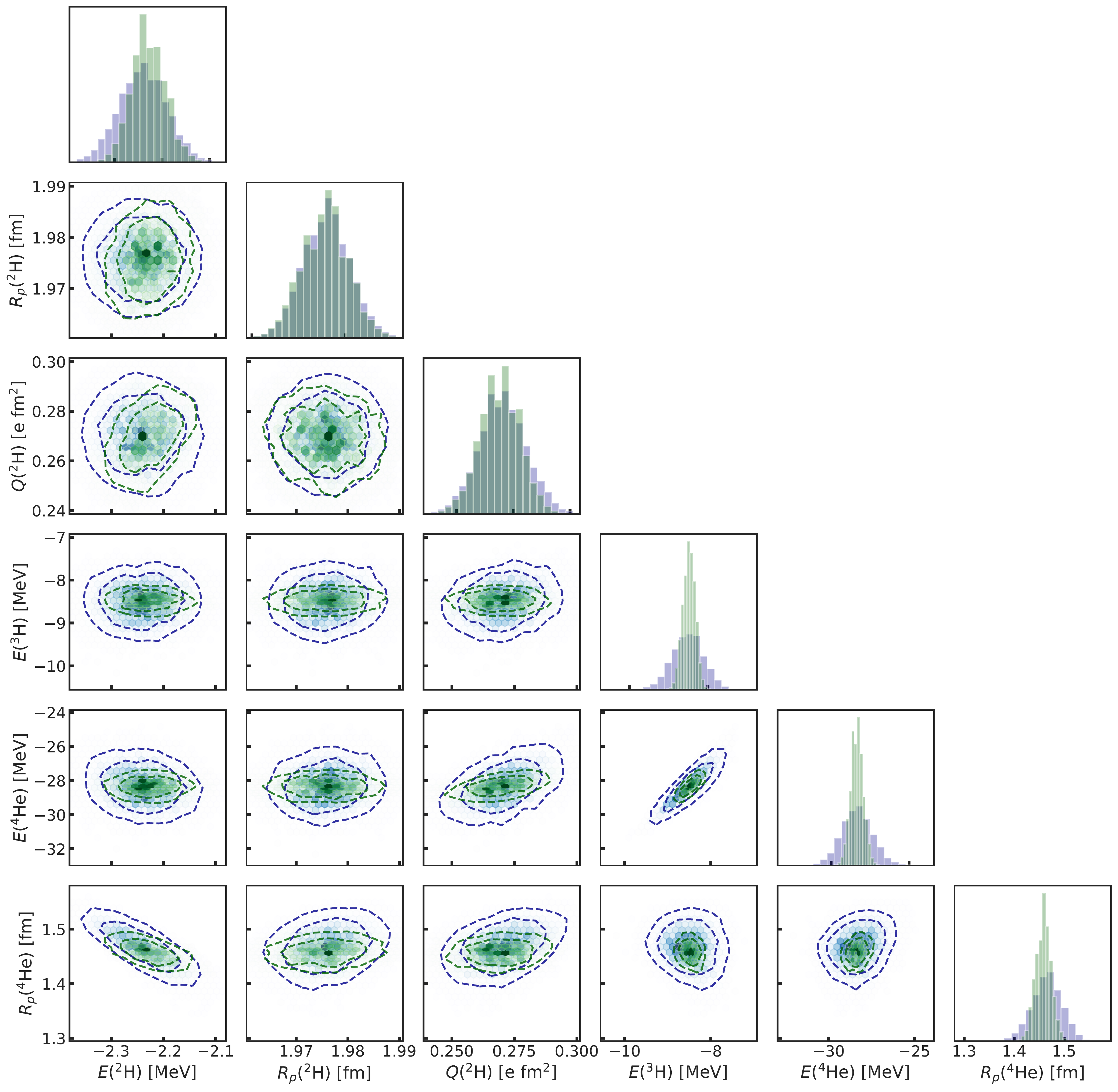}
  \caption{The posterior predictive distribution from sampling over two different posterior distributions. \ppdfew{} (blue) is calibrated by the deuteron observables while \ppdmany{} (green) is calibrated by the deuteron, $^{3}\mathrm{H}$ and $^{4}\mathrm{He}$ observables. The marginal distributions of the observables are shown in the diagonal panels.
 \label{fig:PPD_update}}
\end{figure}

\subsection{Posterior probability updates}
As mentioned in the introduction, the \SIR{} method requires a minimum amount of computation to produce new samples when the posterior \PDF{} is updated for various reasons. Here we present one likely scenario where the posterior is changed due to different choices of calibration data (for instance the inclusion of newly-accessible observables). Let us start from the previously described calibration of our interaction model with three selected deuteron observables. If we add \nuc{3}{H} and \nuc{4}{He} observables into the calibration (experimental target values and error assignments as in Table~\ref{tab:error_assignments}) to further condition the model, the likelihood function needs to be updated accordingly. The sampling of the posterior \PDF{} should be repeated from the beginning and the new samples should be used to construct \PPD{s}. However, using \SIR{} we resample from the same set of prior samples---only with different importance weights. The same set of samples also appear in the sampling of \PPD{s}.
To distinguish the original and the updated posteriors we use the notation \ppdfew{} to denote predictions with only deuteron observable as calibration data and \ppdmany{} with $^{3}\mathrm{H}$ and $^{4}\mathrm{He}$ added to the likelihood. These two different \PPD{s}, generated by \SIR{}, are shown in Fig.~\ref{fig:PPD_update}. Note that the \ppdfew{} (blue) is the same as in Fig.~\ref{fig:PPD}, and is shown here as a benchmark. As expected we observe that the description of $^{3}\mathrm{H}$ and $^{4}\mathrm{He}$ observables is more accurate and more precise  (smaller variations) with \ppdmany{} (green) as compared with \ppdfew{} (blue). We also find that the deuteron ground state energy is slightly improved with the updated posterior. This can be explained by the anti-correlation between $R_\mathrm{p}(^{4}\mathrm{He})$ and $E(^{2}\mathrm{H})$. The additional constraints imposed by $R_\mathrm{p}(^{4}\mathrm{He})$ through the likelihood function propagates to $E(^{2}\mathrm{H})$ via the correlation structure.

\begin{figure}[htbp]
  \includegraphics[width=0.90\textwidth] {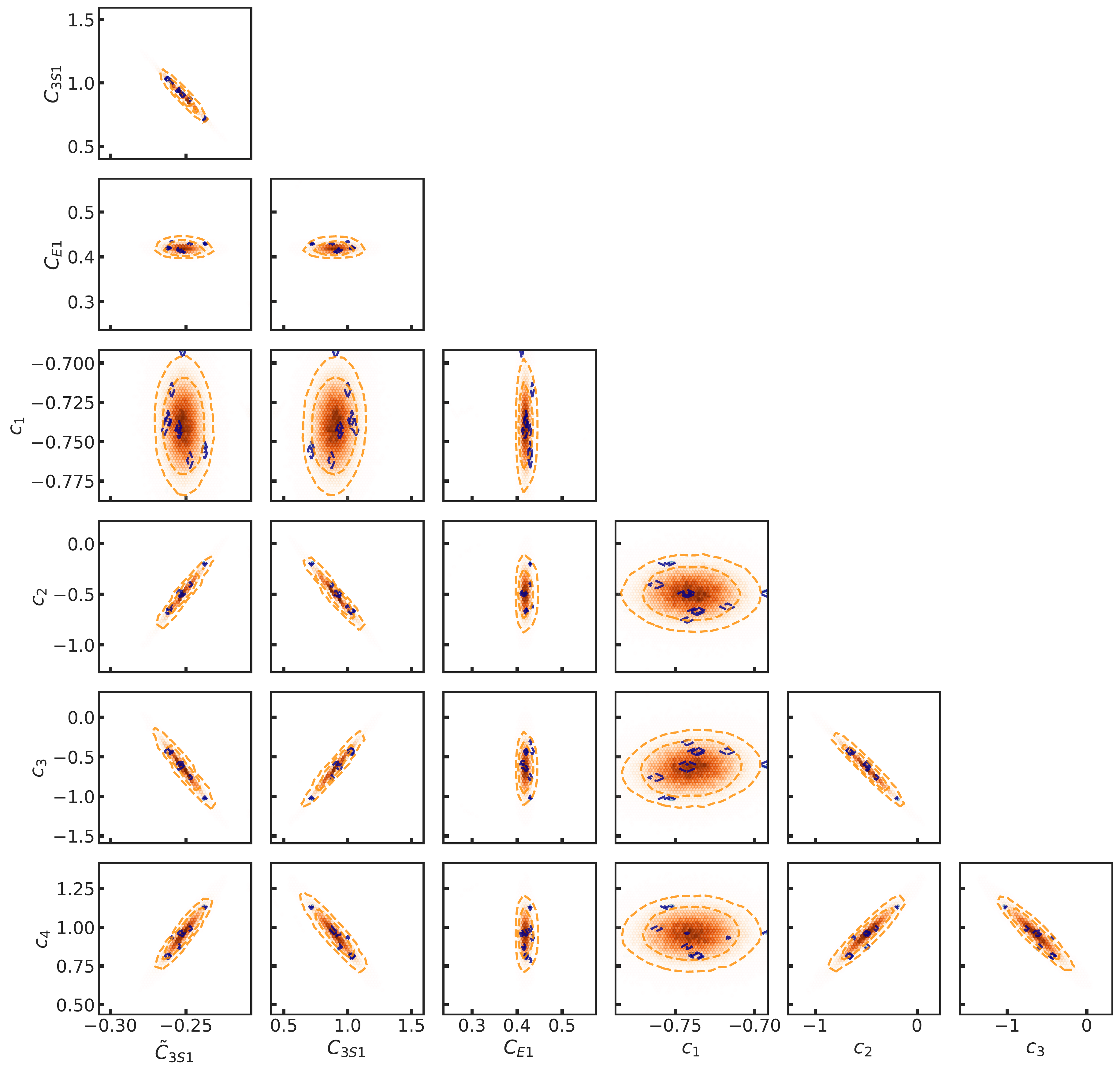}
  \caption{The posterior of LECs sampled with \SIR{} (blue) compared with \MCMC{} sampling (orange) for a situation when the deuteron calibration observables are associated with errors that have been reduced by an order of magnitude (see text for details). The LECs are shown in units of $10^4\, \rm{GeV^{-1}}$,$10^4\, \rm{GeV^{-2}}$ and $10^4\, \rm{GeV^{-4}}$ for $c_{i}$, $\tilde{C}_i$ and $C_i$, respectively. %
    \label{fig:PDF_broken}}
\end{figure}

\begin{figure}[htbp]
  \includegraphics[width=0.90\textwidth] {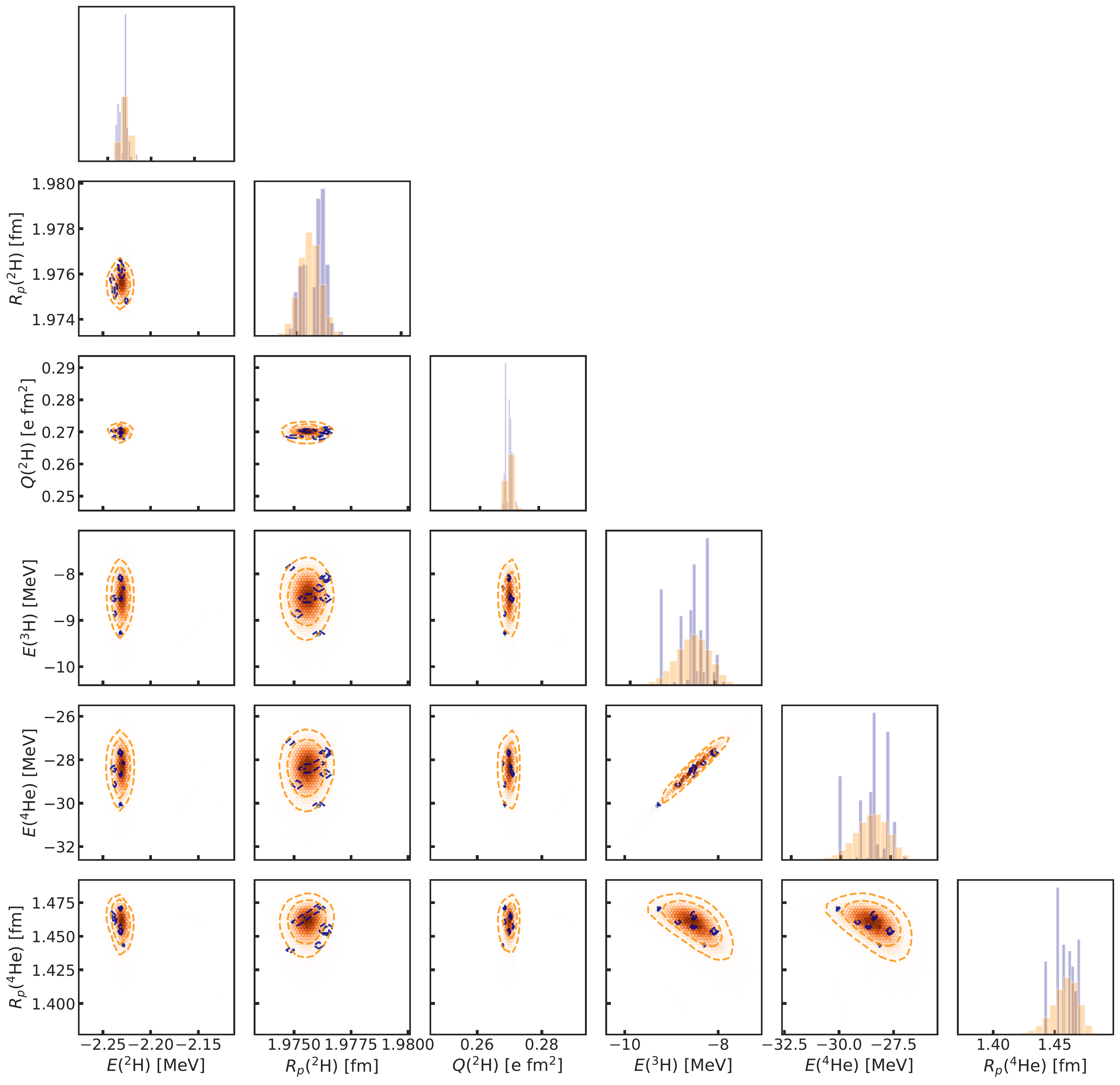}
  \caption{The \PPD{} generated using \SIR{} (blue) and \MCMC{} sampling (orange) for the posterior distributions shown in in Fig.~\ref{fig:PDF_broken}. Marginal histograms of the observables are shown in the diagonal panels.
    \label{fig:PPD_broken}}
\end{figure}

\subsection{\SIR{} limitations}
So far we have focused on the feasibility and advantage of the \SIR{} approach. However, there are some important limitations and we recommend users to be mindful of the number of effective samples. In Fig.~\ref{fig:PPD_update}, we found that our \SIR{} sampling of \ppdfew{} has $n_\mathrm{eff} = 1589.9$, while for \ppdmany{} it drops to $n_\mathrm{eff}  = 314.9$. This can be understood by the resampling from a fixed set of prior samples. The more complex the likelihood function, the less effective the samples. As seen in Fig.~\ref{fig:PPD_update}, the contour lines of \ppdmany{} is less smooth then those obtained from \ppdfew{} due to the smaller number of effective samples. The \SIR{} method will eventually break when $n_\mathrm{eff}$ becomes too small. An intermediate remedy could be the use of kernel density estimators, although that approach typically introduces an undesired sensitivity to the choice of kernel widths.

A similar situation occurs when the target observables are characterized by very small error assignments. This leads to a sharply peaked likelihood function and a decreased overlap with the prior samples. The resulting large variance of importance weights implies that the final set representing the posterior distribution will be dominated by a very small number of samples. 
Here we show such an example where resampling no longer works. We attempted to reconstruct a \PPD{} with only deuteron observables in the calibration, but where all error assignments in Table~\ref{tab:error_assignments} had been reduced by an order of magnitude. The results of this analysis are shown in Figs.~\ref{fig:PDF_broken}, \ref{fig:PPD_broken} which display the \PDF{s} and \PPD{s}, respectively, generated by \SIR{} (blue) and compared with \MCMC{} (orange). The \SIR{} method does not perform well in this case. With $n_\mathrm{eff}=4.4$ the \PDF{} and \PPD{} generated by \SIR{} are represented by a few samples. The \MCMC{} sampling, on the other hand, does manage to identify the updated distribution. 

Unfortunately one can also envision more difficult scenarios in which \SIR{} could fail without any clear signatures. For example, if the prior has a very small overlap with the posterior there is a risk that many prior samples get a similar importance weight (such that the number of effective samples is large) but that one has missed the most interesting region. Again, history matching is a very useful tool in the analysis as it can be used to ensure that we are focusing on the \LEC[s] domain that covers the mode(s) of the posterior.

\section{Summary}
In this paper we reviewed an established sampling method known as \SIR{}. Specifically, we applied importance resampling using the weighted bootstrap algorithm and sampled the posterior \PDF{} for selected \LEC[s] of the $\Delta$NNLO interaction model conditioned on deuteron observables. The resulting \PDF{} and \PPD{} were compared with those obtained from \MCMC{} sampling and a very good agreement was found. We also demonstrated Bayesian updating using \SIR{} by the addition of $^{3}\mathrm{H}$ and $^{4}\mathrm{He}$ observables to the calibration data set. As expected, the predictions of $^{3}\mathrm{H}$ and $^{4}\mathrm{He}$ observables were improved, but also the description of the deuteron ground-state energy which could be explained by the correlation structure between $E(^{2}\mathrm{H})$ and $R_p(^{4}\mathrm{He})$. Finally, we illustrated some limitations of the \SIR{} method that were signaled by small numbers of effective samples. We found that such situations occured when the likelihood became too complex for the limited model, or when prior samples failed to resolve a very peaked posterior that resulted from small tolerances. We also argued that prior knowledge of the posterior landscape is very useful to avoid possible failure scenarios that might not be signaled by the number of effective samples.

\section*{Conflict of Interest Statement}
The authors declare that the research was conducted in the absence of any commercial or financial relationships that could be construed as a potential conflict of interest.

\section*{Author Contributions}
W.G.J. and C.F. contributed equally in this paper.

\section*{Acknowledgments}
This work was supported by the European Research Council under the European Unions Horizon 2020 research and innovation program (Grant No.\ 758027) and the Swedish Research Council (Grant No.\ 2017-04234 and 2021-04507). The computations and data handling were enabled by resources provided by the Swedish National Infrastructure for Computing (SNIC) at Chalmers Centre for Computational Science and Engineering (C3SE), and the National Supercomputer Centre (NSC) partially funded by the Swedish Research Council through Grant No.\ 2018-05973.

\bibliography{./master}

\end{document}